\documentclass[11pt,twoside]{article}

\usepackage{asp2006}
\usepackage{epsf}

\begin{document}
\title{The Environmental Dependencies of Star-formation and the Origin
of the Bimodality in Galaxy Properties}   
\author{C. P. Haines, A. Gargiulo, A. Mercurio, P. Merluzzi, F. La
  Barbera, \\G. Busarello, \& M. Capaccioli}   
\affil{INAF Osservatorio Astronomico di Capodimonte, via Moiariello
  16, I-80131 Naples, Italy}    

\begin{abstract} 
We examine the origins of the bimodality observed in the global
properties of galaxies by comparing the environmental dependencies of
star-formation for giant and dwarf galaxy populations. Using Sloan
Digital Sky Survey (SDSS) DR4
spectroscopic data to create a volume-limited sample complete to M$^{\star}+3$,
we find that the environmental dependences of giant
and dwarf galaxies are quite different, implying fundamental
differences in their evolution.
Whereas the star-formation histories of
giant galaxies are determined primarily by their merger history,
resulting in passively-evolving giant galaxies being found in all
environments, we show that this is not the case for dwarf galaxies. In
particular, we find that old or passive dwarf galaxies are {\em only} found as
 satellites within massive halos (clusters, groups
or giant galaxies), with none in the lowest density regions. This
implies that star-formation in dwarf galaxies must be much more
resilient to the effects of mergers, and that the evolution of dwarf
galaxies is primarily driven by the mass of their host halo, through
effects such as suffocation, ram-pressure stripping or galaxy
harassment.
\end{abstract}


\section{Introduction}

The global properties of galaxies have been found to be bimodally
distributed about a stellar mass $\sim3\times10^{10}{\rm M}_{\sun}$
(M$^{\star}+1$), with more massive galaxies predominately passive, red
spheroids dominated by old stellar populations, and less massive
galaxies tending to be blue, star-forming disk galaxies whose light is
dominated by young stars \citep[e.g.][]{kauffmann03}. This implies
fundamental differences in the formation and evolution of giant and
dwarf galaxies. 
 What causes this bimodality\,? If galaxies grow
hierarchically through merging and accretion, why do they only become passive once they reach
$\sim3\times10^{10}{\rm M}_{\sun}$\,?

One approach to this problem is to look at the environmental trends of
galaxies, as the trends with mass are paralleled by those with
environment. In particular, 
passively-evolving spheroids dominate cluster cores whereas galaxies
in field regions are typically star-forming disk galaxies, giving rise
to the classic morphology-density and star formation-density relations. 
Are these environmental trends : (i) the direct result of the initial
conditions in which the galaxy forms, whereby cluster galaxies could form
earlier and evolve more rapidly through a more active merger history,
than those in the smoother lower density regions; or (ii) produced later by the
direct interaction of galaxies with one or more aspects of their
environment through processes such as galaxy harassment, suffocation
or ram-pressure stripping\,? This is the so-called nature versus
nurture problem. 

Studies of the most massive galaxies find little or
no difference in the mass-to-light ratios or mean stellar ages between
cluster and field early-type galaxies, implying that environmental
processes are not important \citep[e.g.][]{vandokkum}. However, at fainter magnitudes large
variations with local density are seen in the ages, colours and the
shape of the luminosity function, implying that environmental process
are much more important for low-mass galaxies \citep[e.g.][]{sos,mercurio}.
By examining when, where and how galaxies are being
transformed, we can gain information as to the nature of the physical
mechanisms responsible for the transformation. By in addition, seeing
how these environmental dependences vary with mass, we can hope to
understand the causes of the bimodality. 

We have examined the origins of the bimodality by comparing the
environmental dependencies of giant and dwarf galaxy populations 
in the vicinity of the supercluster centred on the rich cluster
A\,2199 at \mbox{$z=0.0309$}. 
This is the richest low-redshift \mbox{($z<0.04$)} structure
covered by SDSS DR4, producing a spectroscopic sample of \mbox{$\sim2$\,000}
galaxies that is $\sim9$0\%
complete to a magnitude limit of \mbox{M$_r=-17.8$} or \mbox{M$^{\star}\!+3.3$}, i.e. well
into the dwarf regime. 
From these we measured global trends with
environment (using the adaptive kernel estimator to estimate the local
galaxy density on scales of the host halo) for both giant \mbox{(M$_r<-20$)}
and dwarf \mbox{($-19<{\rm M}_r<-17.8$)} subsamples using the $r$-band
luminosity-weighted mean stellar age and H$\alpha$ emission as two
independent measures of star-formation history \citep[for details see][]{haines}. 

\section{Results}

One example of the global bimodality in galaxy properties is seen in 
the relation between mean stellar age and the $r$-band absolute
magnitude (M$_r$), with a
population of bright \mbox{($\sim\!L^{\star}$)} galaxies \mbox{$\sim\!10$\,Gyr} old,
and a distinct population of fainter galaxies dominated by young
\mbox{($<3$\,Gyr)} stars \citep[see Fig. 1 of][]{haines}. 
Examining the environmental trends of mean stellar age, we find that
both giant and dwarf galaxy populations get steadily older with
density, and in all environments giant galaxies are at least 1\,Gyr
older on average than dwarf galaxies. In high-density regions
corresponding to cluster cores, galaxies are predominately old
\mbox{($\ga\!8$\,Gyr)}, independent of their luminosity. Moreover, while the
age distribution of massive galaxies extends to include ever younger
ages with decreasing density, that of dwarf galaxies gets younger but
also narrows, so that at the lowest-densities found in the rarefied
field only young \mbox{($\la\!2$\,Gyr)} galaxies are found. 
Equally, examining the environmental dependence of the fraction
of old \mbox{($>\!7$\,Gyr)} galaxies, we find
that in the highest density regions, \mbox{$\sim7$5\%} of both giant and
dwarf galaxies are old. Whereas the fraction of giant galaxies with
old stellar populations declines gradually with decreasing density to
the global field value of \mbox{$\sim5$0\%}, that of dwarf galaxies drops
rapidly to \mbox{$\sim2$0\%} by the cluster virial radius, and continues to
decrease, tending to {\em zero} for the lowest density bins. Identical
trends are independently observed when considering passive galaxies
with \mbox{EW(H$\alpha)<4$\AA}.     
  
\begin{figure}
\plotone{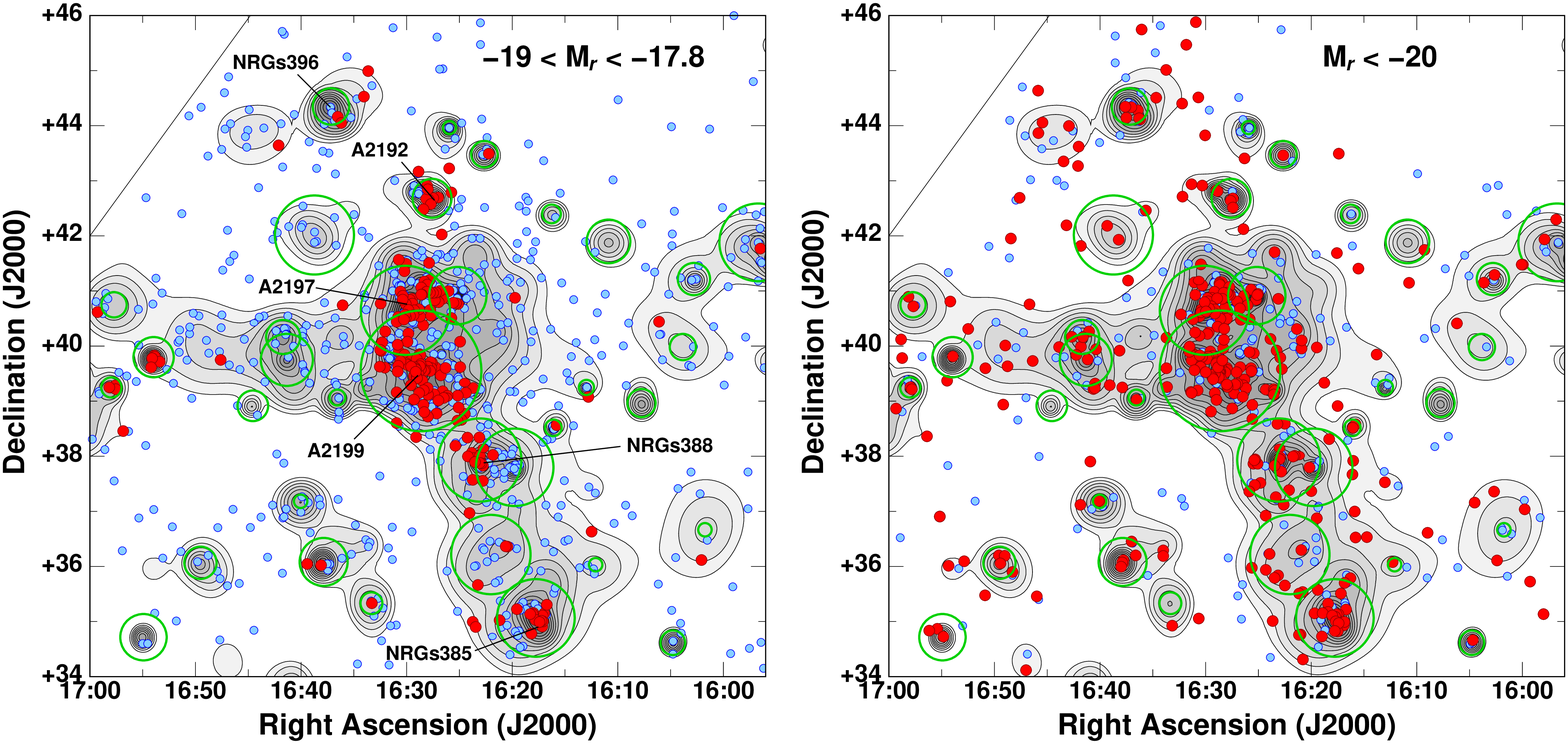}
\caption{The distribution of galaxies with (light-blue circles) and
  without (\mbox{EW[H$\alpha]\leq4$\AA}; red circles) H$\alpha$
  emission in the A2199 supercluster environment, for dwarf (\mbox{$-19<{\rm
  M}_{r}<-18$}; left) and giant (\mbox{M$_{r}<-20$}; right) galaxies. The
  black contours represent the local luminosity-weighted surface
  density of galaxies with redshifts within \mbox{2\,000\,km\,s$^{-1}$} of
  A2199. The
  green circles indicate the virial radii of the galaxy
  groups/clusters associated with the A2199 supercluster.}
\label{spatial}
\end{figure}

To relate the differences in environmental trends directly to the
effect of the supercluster, Figure~\ref{spatial} shows the spatial
distribution of passively-evolving (red solid circles) and
star-forming (light-blue circles) galaxies, for the dwarf (\mbox{$-19<{\rm
  M}_{r}<-17.8$;} left panel) and giant (\mbox{M$_{r}<-20$;} right panel)
galaxy subsamples, with relation to the supercluster as represented by the
grey-scale isodensity contours. 
Although the passively-evolving giant galaxies are
more contrated towards the centres of the rich clusters in the
supercluster than their star-forming counterparts, they are found
throughout the region covered. In the low-density regions there is an
equal interspersed mixture of passive/old and star-forming/young
galaxies. This indicates that their evolution is driven primarily by
internal mechanisms and their merging history rather than by direct
interactions with their large-scale environment; the gradual overall
trends of star-formation with environment reflect the increasing
probability with density that a galaxy with have undergone a major
merger during its lifetime. 

In contrast, the
star-formation history of dwarf galaxies are strongly correlated with
their environment. While the cores of clusters are still dominated by
passively-evolving dwarf galaxies, elsewhere almost all of the dwarf
galaxies are currently actively star-forming, and of the few remaining
passively-evolving galaxies outside a cluster, {\em all} are found
in either poor groups or within $\sim20$0\,kpc of an old, massive
galaxy. None are found to be isolated.
 
Extending the study to the entire SDSS DR4 dataset, we consider a
volume-limited sample of \mbox{$\sim3$0\,000} galaxies in the redshift range
\mbox{$0.005<z<0.037$}, complete to \mbox{M$_{r}=-18.0$}. Examining the
fraction of passively-evolving galaxies as a function of both their
luminosity/stellar mass and local environment, we find that in
high-density regions passively-evolving galaxies dominate independent
of luminosity, making up \mbox{$\approx\!7$0\%} of the population. In the rarefied
field however, the fraction of passively-evolving galaxies is a strong
function of luminosity, dropping from 50\% for \mbox{M$_{r}\la\!-21$}, to
{\em zero} by \mbox{M$_{r}=-18$} or a stellar mass \mbox{$\sim\!10^{9.2}{\rm M}_{\sun}$}. Indeed, in the lowest-luminosity range
covered \mbox{($-18<{\rm M}_{r}<-16$)} {\em none} of the \mbox{$\simeq6$00} galaxies
in the lowest density quartile are passive. 
These results confirm and extend the
well known observation that dwarf ellipticals in the local Group are found only
near massives galaxies \citep[e.g.][]{binggeli,ferguson}, or in more massive structures such as the
Virgo and Coma clusters \citep[e.g.][]{conselice}, and  
 imply that processes {\em internal} to the galaxy cannot completely
shut down star-formation in dwarf galaxies, and instead they only become
passive once they become a satellite within a more massive halo.


\section{Discussion}

The relationship between star-formation and environment for giant and
dwarf galaxies are quite different. Whereas the star-formation
histories of giant galaxies are primarily determined by their merger
history, star-formation in dwarf galaxies appears much more resilient
to the effects of mergers. Instead dwarf galaxies become passive only
once they become satellites within a more massive halo, by losing
their halo gas reservoir to the host halo (``suffocation'') or other
environment-related processes such as galaxy harassment and/or ram-pressure
stripping. These differences can be understood in the context of the
hot and cold gas infall \citep{dekel} or AGN feedback models of galaxy
evolution. When two massive gas-rich galaxies merge, tidal forces
trigger a star-burst and fuel the rapid growth of the central black
hole, until outflows from the AGN drive out the remaining cold gas
from the galaxy, rapidly terminating the starburst \citep{springel}. In massive
galaxies, the halo gas is also heated by stable virial shocks, and is
prevented from cooling by feedback from the quiescent accretion of the
hot gas onto the black hole, effectively permanently shutting down
star-formation \citep{croton}. Because black-hole growth is strongly dependent on
galaxy mass, AGN feedback in low-mass galaxies is much less efficient
at expelling cold gas or affecting star-formation. Low-mass galaxies
also regulate their star-formation through supernovae feedback
preventing starbursts that exhaust the cold gas, which is in turn
constantly replenished by filamentary cold streams from the halo, so that their
ability to maintain star-formation over many Gyr is much less affected
by their merger history. Thus the characteristic mass-scale of the
bimodality represents the point at which galaxies can become passive
through internal processes.



\end{document}